\documentclass[twocolumn,showpacs,showkeys,preprintnumbers,amsmath,amssymb]{revtex4}

\usepackage{graphicx}
\usepackage{dcolumn}
\usepackage{bm}

\begin{document}


\title{Slow failure of quasi-brittle solids}

\author{Leonid S.Metlov}
\email{lsmet@kinetic.ac.donetsk.ua}
\affiliation{Donetsk Institute of Physics and Engineering, Ukrainian
Academy of Sciences,
\\83114, R.Luxemburg str. 72, Donetsk, Ukraine}

\thanks{I thank all my colleagues who put me hard questions that essentially develop this approach}

\date{\today}

\begin{abstract}
A new mesoscopic non-equilibrium thermodynamic approach is developed. The approach is based on the thermodynamic identity associated the first and second law of thermodynamics. In the framework of the approach different internal dissipative channels of energy are taken in account in an explicit form, namely, the thermal channel and channels of defect subsystems. The identity has a perfect differential form what permits to introduce an extended non-equilibrium state and use the good developed mathematical formalism of equilibrium and non-equilibrium thermodynamics. The evolution of non-equilibrium variables of a physical system are described by a Landau-based equation set expressed through internal or different kinds of free energy connected by means of the Legendre transforms. The accordance between the different kinds of energy is possible owing to introduction of some trends into the equation subset described the defect subsystems and having a nature of structural viscosity. The possibilities of the approach are illustrated on the example of quasibrittle solid damage and failure. Taking into account only one type of defects (viz., microcracks) and mechanical parameters in an expansion of free energy down to third powers in relative to average energy per microcrack, the description of destruction of quasi-brittle solids during long-term loading is considered. The consideration allows to find equilibrium and non-equilibrium values of the free energy. A qualitative behavior of the system on parameters of the theory is analyzed. The destruction of material is described from the uniform positions, both at uniform tension and uniaxial compression. Origins of the high stability of mine workings at small depths and their instability at large depths are explained.
\end{abstract}


\maketitle

\section{Introduction}

A problem of solid fracture is one of the most difficult challenges for modern science. The fracture is a compound multilevel process that depends on many factors, such as physical properties of material, its structural organization and environment and so on. Variety of materials and environments predetermines many fracture mechanisms that are distributed in the range from brittle to plastic failure. These are sudden dynamic destruction during quick loading and a slow accumulation of microdefects with their subsequent coalescence during a long-term static loading.

Rocks are one of types of poorly studied solids of peculiar structural organization. Rock properties to resist to long-term destructive effects of rock pressure at large depths are important component of stabilization of underground structures. At the same time rocks demonstrate a number of common for all solids features of behavior that can be of certain interest for investigation of processes that are initiated by a strong external effect on a matter and also for investigation of the resulting non-equilibrium states.

The governing relaxation mechanism in the case of quasi-brittle solids is accumulation and growth of microcracks, its coalescence and branching and finally destruction of a solid \cite{CO96}, \cite{RBN05}. Previously this mechanism for the case of large deformation was considered of general thermodynamic positions \cite{VV01}, \cite{Mish97}, \cite{Met97}, \cite{Met02}, \cite{Met07}. Applicability of equilibrium thermodynamics was limited by intermediate time scale less than that of necessary for diffusive healing of microcracks. It was expected that a real solid has already natural weak points, non-continuities and so on, which became more intense due to external force action.

Rock was considered usually as passive mechanical continuum, properties of which are evolved just under the influence of active external stresses. Yet, it is clear from general considerations that in the process of metamorphism at large depths the rocks reach state what is equilibrium in thermodynamic meaning for the given depths. In these conditions, the formation of an artificial cavity results not only to mechanical disequilibrium, but, what is much more essential, to thermodynamic disequilibrium of rocks. Actually rocks turn out in a non-equilibrium state and relax by means of microcrack accumulation, subsequent coalescence of them and macroscopic destruction. Depending on the level of accumulated internal energy, the autowave processes of different kinds, which are typical for an active media, are possible \cite{KO94}, including generation of circular zonal destruction structures around mine workings \cite{Shem86}, \cite{MMZ02}.

In this work on the basis of the principles of thermodynamics an attempt to formulate a consistent theory is made that takes into account both mentioned factors – energy accumulated in the form of internal stresses and work of external stresses. Areas of increased internal stresses make additional contribution to internal energy; generation of microcracks also results in growth of internal energy of a solid of the external work which is equal to a total energy of break bonds.

\section{DEFINING RELATIONS}

For infinitely small changes of external parameters of a defect-free homogeneous system during its interaction with external bodies a change of internal energy (first law of thermodynamics) is equal to
  \begin{equation}\label{a1}
du=T\delta s+\sigma_{ij} d\varepsilon_{ij}^{e} +\sigma_{ij} d\delta \varepsilon_{ij}^{n},
  \end{equation}
 where $u$ is density of internal energy, $J/ m^{3}$; $T$ is temperature, $K$; 
  $s$ is entropy density (here from external sources only), $J/K*m^{3}$; 
$\sigma_{ij}$ is the stress tensor, $H /m^{2}$ and $\varepsilon_{ij}^{e}$, $\varepsilon_{ij}^{n}$
are the reversible and non- reversible parts of deformation tensor. Here increments of variables, which are state functions, are perfect differentials. In this form the law expresses external forced and thermal actions on the studied object and it fits for description of both equilibrium and non-equilibrium processes.

Irreversible part of the external work goes to internal heat at the cost of entropy production by internal sources and to generation and modification of defect subsystem, what may be written as
  \begin{equation}\label{a2}
\sigma_{ij} \delta \varepsilon_{ij}^{n}=T\delta s''+\chi \delta s' + 
\sum_{k=1}^{N}\varphi_{k} \delta h_{k}\geq 0.
  \end{equation}
Here $s''$ is that part of entropy produced by internal sources, which has time to go to equilibrium state during the external act, $s'$ is that part of entropy produced by internal sources, which remains yet into the non-equilibrium state and it lies ahead to relax in the next time, $\chi$ is the non-equilibrium temperature, $s\varphi_{k}$, $h_{k}$ are average $k$-type defect energy ($J$) and defect density ($m^{-3}$), respectively, $N$ is number of defect subsystems. The relation (\ref{a2}) is one of equivalent form of the second law of thermodynamics. Substituting it to (\ref{a1}) and associating equilibrium entropy produced by external and internal sources one may deduce next identity
  \begin{equation}\label{a3}
du=T d s+\sigma_{ij} d\varepsilon_{ij}^{e} +\chi ds'+
\sum_{k=1}^{N}\varphi_{k} d h_{k}.
  \end{equation}

The relation (\ref{a3}) is combination of the first and second law of thermodynamics in the form of identity, but not in the known inequality form. It turns out possible to write it in this form as all internal channels of energy dissipation are known for the given model of a solid. The sound property of this identity is that all increments of the thermodynamic variables are perfect differentials in one or another sense. The first two terms are treated in the sense of equilibrium thermodynamics; the second two terms describe non-equilibrium processes passing in a solid. The temperature and stress tensor may be calculated by means of differentiation of internal energy with respect to full equilibrium entropy and elastic strain tensor as it does in the classical thermodynamics. The non-equilibrium temperature and defect energy are too calculated by means of differentiation of internal energy with respect to non-equilibrium entropy and defect densities
\begin{equation}\label{a4}
    \chi=\dfrac{\partial u}{\partial s'},
    \quad \varphi_{k}= \dfrac{\partial u}{\partial h_{k}},
\end{equation}
but in the sense of the extended thermodynamic forces in the Landau-based equations. Owing to this, the expression (\ref{a3}) may be regarded as complete differential with respect to variables $s$, $\varepsilon_{ij}^{e}$, $s'$, and $h_{k}$ and this variables determines some generalized non-equilibrium state of the system. What does it give?  Now one can use the simple mathematics of equilibrium thermodynamics for the non-equilibrium case. Time evolution of non-equilibrium thermodynamic parameters $s'$, and $h_{k}$ can be calculated by means of set of equations each of it is Landau-Chalatnikov or Ginzburg-Landau equation type \cite{Met07}, \cite{LP86}:
\begin{eqnarray}\label{a5}
\nonumber
\dfrac{\partial s'}{\partial t}=\gamma_{s'} \dfrac{\partial u}{\partial s'}=
\gamma_{s'} \chi,  \\
 \dfrac{\partial h_{k}}{\partial t}=\gamma_{h_{k}} (\dfrac{\partial u}{\partial h_{k}}
-\overline{\varphi}_{k})=\gamma_{h_{k}} (\varphi_{k}-\overline{\varphi}_{k}).
\end{eqnarray}

These equations differ from the classical prototype. First, here the internal energy is used as a more fundamental quantity entered to the first law of thermodynamics. Second, in the right parts of equations the sign «plus» stands it means that the stable stationary states are placed near a maximum of the internal energy. This peculiarity is conditioned by energetic pumping in the cost of irreversible work. The third difference is presence of viscosity shift $\overline{\varphi}_{k}$ in the equation describing defect subsystem. In consequence of the shift the stable stationary states do not coincide exactly with the maximums of the internal energy, but coincide with some points in which the curve of the internal energy has finite positive slope of tangents. This shift has same physical sense as order parameter in the case of a supercooled liquid and plays an important role for construction of a consistent set of non-equilibrium thermodynamic potentials connected by means of the Legendre transformations.

So as number of conjugate pairs of thermodynamic variables increases in comparison with equilibrium case the number of the Legendre transformations increases too. For example, if one carries out the Legendre transformation $f=u-T s$ the «classic» free energy is resulted. Difference of it from classic one is that this type of free energy has non-equilibrium nature. The stable stationary states of a system in reference to defect subsystem are placed near the maximums of the free energy same as for internal energy.

If one carries out the Legendre transformation in the form
  \begin{equation}\label{a6}
\tilde{f}=u-T s-\sum_{k=1}^{N}\varphi_{k} h_{k}
  \end{equation}
the complete free energy is resulted. Despite to the above the stable stationary states of the system in reference to defect subsystem are placed near the minimums of this free energy. Its total differential can be written as
  \begin{equation}\label{a7}
d\tilde{f}=-sdT+\sigma_{ij} d\varepsilon_{ij}^{e}+\chi ds'-\sum_{k=1}^{N} h_{k}d \varphi_{k} .
  \end{equation}

From this expression one notices that the complete free energy is function of such own arguments as $T$, $\varepsilon_{ij}^{e}$, $s'$ and $\varphi_{k}$. Time evolution of non-equilibrium thermodynamic parameters $s'$ and $\varphi_{k}$ can be calculated by means of set of equations which is similar to (\ref{a5})
\begin{eqnarray}\label{a8}
\nonumber
\dfrac{\partial s'}{\partial t}=\gamma_{s'} \dfrac{\partial \tilde{f}}{\partial s'}=
\gamma_{s'} \chi,  \\
 \dfrac{\partial \varphi_{k}}{\partial t}=
\gamma_{\varphi_{k}} (\dfrac{\partial \tilde{f}}{\partial \varphi_{k}}
-\overline{h}_{k})=\gamma_{\varphi_{k}} (h_{k}-\overline{h}_{k}).
\end{eqnarray}

The non-equilibrium temperature and defect density are calculated by means of differentiation of the complete free energy with respect to non-equilibrium entropy and defect energy
\begin{equation}\label{a9}
    \chi=\dfrac{\partial \tilde{f}}{\partial s'},
    \quad h_{k}= -\dfrac{\partial \tilde{f}}{\partial \varphi_{k}}.
\end{equation}

These relations and relations (\ref{a4}) can be interpreted as the state equations. Ignoring the thermal effects and supposing that the drive parameter $\varepsilon_{ij}^{e}$  is a constant the state equation can be written in two equivalent forms:
\begin{equation}\label{a10}
    \varphi_{k}= \dfrac{\partial u}{\partial h_{k}},
    \quad h_{k}= -\dfrac{\partial \tilde{f}}{\partial \varphi_{k}}.
\end{equation}

\section{POWER SERIES ANALYSIS}

Viewing only one type of defect power series expansions of the internal and free energies in terms of its own arguments can be written in the above approximation as
  \begin{equation}\label{a11}
u=u_{0}+\varphi_{0} h-\frac{1}{2} \varphi_{1} h^{2}+\frac{1}{3} \varphi_{2} h^{3}-
\frac{1}{4} \varphi_{3} h^{4} ,
  \end{equation}
  \begin{equation}\label{a12}
\tilde{f}=\tilde{f}_{0}-h_{0} \varphi+\frac{1}{2} h_{1} \varphi^{2}-\frac{1}{3}h_{2} \varphi^{3}+
\frac{1}{4}h_{3} \varphi^{4}  .
  \end{equation}

For simplicity sake, only one type of defects is considered here. The coefficients $\varphi_{0}$, $\varphi_{1}$, $\varphi_{2}$ and $\varphi_{3}$ characterize energetic processes at microscopic levels, the coefficients $h_{0}$, $h_{1}$, $h_{2}$ and $h_{3}$ characterize dependence of defect density from processes at microscopic levels. Different powers in the expansions describe thermodynamic processes of different levels. The main process is described by the lowest power; the processes corrected to it are described by higher powers of the expansions. Alternation of the signs is due to the Le Chatelier principle – the processes of each next level point in opposite direction to the processes of the previous level. The «state equations» for each case have the form:
  \begin{equation}\label{a13}
\varphi=\dfrac{\partial u}{\partial h}=\varphi_{0}-\varphi_{1} h+\varphi_{2} h^{2}-\varphi_{3} h^{3} ,
  \end{equation}
  \begin{equation}\label{a14}
h=-\dfrac{\partial \tilde{f}}{\partial \varphi}=h_{0}-h_{1} \varphi+h_{2} \varphi^{2}-h_{3} \varphi^{3}  .
  \end{equation}

Let us notice that in the linear (quadratic for energy) approximation the dependence between thermodynamic variables $\varphi$ and $h$ is biunique in the both cases. Taking in account that in this approximation the both equations must lead to the same physical result the connection between expansion coefficients is next:
\begin{equation}\label{a15}
    \varphi_{0}= \dfrac{h_{0}}{h_{1}},
    \quad \varphi_{1}= \dfrac{1}{h_{1}},
\end{equation}
\begin{equation}\label{a16}
    h_{0}= \dfrac{\varphi_{0}}{\varphi_{1}},
    \quad h_{1}= \dfrac{1}{\varphi_{1}}.
\end{equation}

Calculation of the internal and free energy and the state equation at model parameters of $u_{0}=0.114\cdot 10^{6}J/m^{3}$, $\varphi_{0}=0.113\cdot 10^{-1}J/m^{2}$, $\varphi_{1}=0.0.976\cdot 10^{-12}J/m$ is presented in the fig. \ref{f1}a. 
\begin{figure*}
  \includegraphics [width=3.0 in] {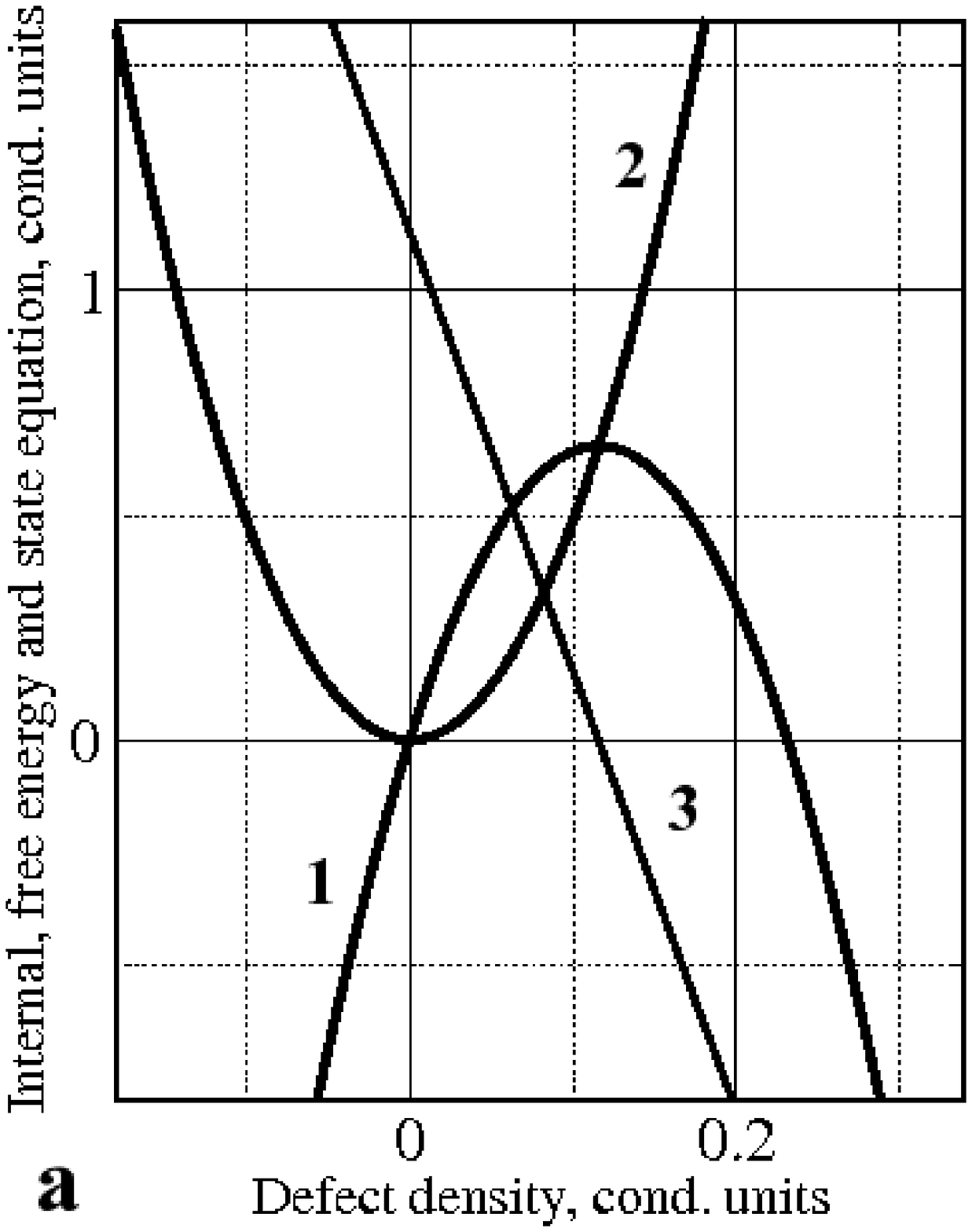}
  \includegraphics [width=3.0 in] {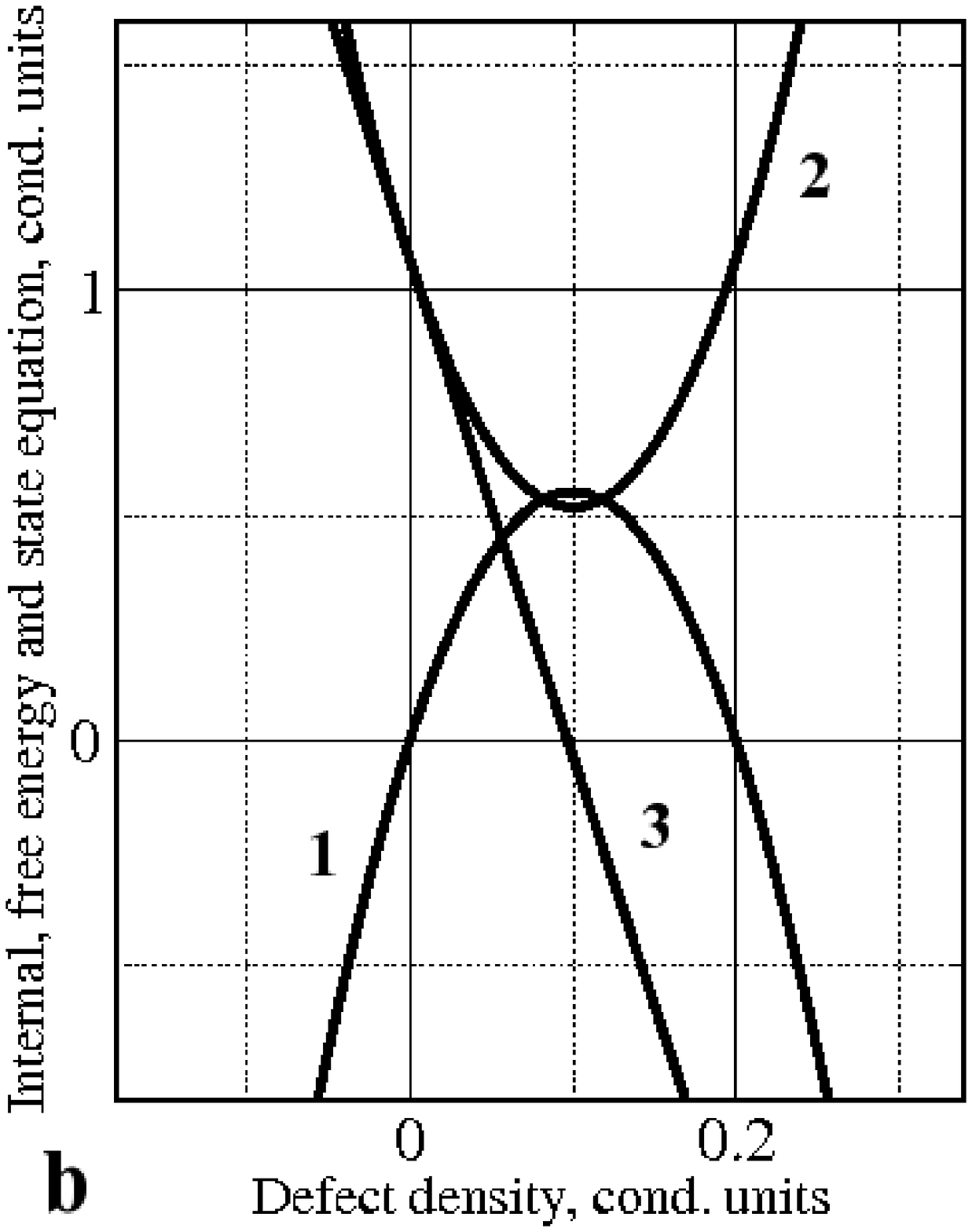}
  \caption{\label{f1} Dependence of the thermodynamic energies versus defect density:  
  a - initial ones; 
  b - shifted ones; 
1 – internal energy; 2 – free energy; 3 – state equation $\varphi=\varphi(h)$}\label{f1}
\end{figure*}
The dimension of the variable $h$ is $m^{-1}$. If in accordance with the state equation (\ref{a13}), (\ref{a14}) written in the linear approximation one makes transition from variable $h$ to variable $\varphi$. then sequence of extremes are changed, the maximum of the internal energy (at zero value of $\varphi$) is placed in the left side, the minimum of free energy is placed in the right side.

From the fig. \ref{f1}a one notices that the maximum of the internal energy does not coincide with the minimum of the free energy. The standard procedure of minimization of free energy has led to the zero value of the variable $h$ as in the same time the analogous procedure of maximization of internal energy has led to the zero value of the variable $\varphi$ suggesting that there is some conflict between descriptions in terms of the internal and free energy. It is possible to eliminate this conflict offsetting the extremums of potentials (energies) to meet one another in such away that they both turn out in the same point satisfying to the state equation. In the terms of the initial potentials (energies) it means the stable point is that in which the slopes of tangents to the internal and free energy curves are not equal to zero and together they are satisfied to the state equation. In fact, an arbitrary point between zero and $h_{max}$ satisfies to this condition, consequently the position of the point must be determined from experimental data.  The plots of shifted thermodynamic potentials for value $\overline{h}=0.1m^{-1}$ are presented in the fig. \ref{f1}b. The value of $\overline{\varphi}$ determined from the state equation is $0.15\cdot Jm^{-1}$.

From the figure we notice that the curve of internal energy is shifted a little to the left side, and the curve of free energy is essentially shifted in the right side so as the extremums of both energies are coincided. Thus, a system during its evolution tends to common stable state independently of what kind of energy is used for description. In consequently of it the evolution of a system can be described by two kinds of Landau-Khalatnikov equations expressed in terms of internal energy (\ref{a5}) or free one (\ref{a8}).

All along, it is possible to determine the kinetic coefficients $\gamma_{h}$ and $\gamma_{\varphi}$ so that the equations (\ref{a5}) and (\ref{a8}) at a arbitrary time will describe one and the same state associated with thermodynamic pair $h$ and $\varphi$ connected by the state equation (\ref{a13}) or (\ref{a14}). The evolution is finished with achievement of one and the same system state associated with pair $\overline{h}$ and $\overline{\varphi}$. 

\section{FAILURE OF QUASIBRITTLE MATERIALS}

\subsection{General}

In a special case of isothermal processes, which is just considered here, valid is
  \begin{equation}\label{a17}
d\tilde{f}=\sigma_{ij} d\varepsilon_{ij}^{e}-\sum_{k=1}^{N} h_{k}d \varphi_{k} ,
  \end{equation}
so $\tilde{f}=\tilde{f}(\varepsilon_{ij}^{e},\varphi_{k})$ If explicit expression of the free energy density dependence of its arguments is known, then variables $\sigma_{ij}$ and $h_{k}$ can be determined by simple differentiation. As determination of the explicit dependence for the density of the free energy is intractable problem, we use a standard method for such cases, expansion of the free energy into a series up to cubic power of its arguments:
\begin{eqnarray}\label{a18}
\nonumber
\tilde{f}(\varepsilon_{ij}^{e},\varphi)=\tilde{f_{0}}-h_{0} \varphi+\frac{1}{2} h_{1} \varphi^{2}-\frac{1}{3}h_{2} \varphi^{3}+ \\
+\frac{1}{2}\lambda (\varepsilon_{ii}^{e})^{2}+\mu (\varepsilon_{ij}^{e})^{2}-
g\varphi\varepsilon_{ii}^{e}+ \\
\nonumber 
+\frac{1}{2}\overline{\lambda}\varphi(\varepsilon_{ii}^{e})^{2}+
\overline{\mu}\varphi(\varepsilon_{ij}^{e})^{2}+e\varphi^{2} \varepsilon_{ii}^{e} .
\end{eqnarray}
Here we consider only one type of defect, which is main for quasibrittle failure, namely, micro-cracks. Then $h$ is micro-crack density and $\varphi$ is average energy per one crack. During an external mechanical action the micro-cracks are produced, collected and merged into macroscopic clusters such as main-cracks and so on.

Corresponding «state equations» have the form:
\begin{eqnarray}\label{a19}
\nonumber
h=h_{0}-h_{1} \varphi+h_{2} \varphi^{2}+g\varepsilon_{ii}^{e}+ \\
\nonumber
+\frac{1}{2}\overline{\lambda}(\varepsilon_{ii}^{e})^{2}+
\overline{\mu}(\varepsilon_{ij}^{e})^{2}-2e\varphi \varepsilon_{ii}^{e}, \\
\sigma_{ij}=\lambda\varepsilon_{ii}\delta_{ij}+2\mu\varepsilon_{ij}-g\varphi\delta_{ij} \\
\nonumber
-\overline{\lambda}\varphi\varepsilon_{ii}\delta_{ij}
- 2\overline{\mu}\varphi\varepsilon_{ij}+e \varphi^{2}\delta_{ij}.
\end{eqnarray}

The evolution equation in an explicit form now is
\begin{eqnarray}\label{a20}
\nonumber
\frac{\partial\varphi}{\partial t}=\chi_{\varphi} (h_{0}-\overline{h}-h_{1} \varphi+h_{2} \varphi^{2}+g\varepsilon_{ii}^{e}+ \\
+\frac{1}{2}\overline{\lambda}(\varepsilon_{ii}^{e})^{2}+
\overline{\mu}(\varepsilon_{ij}^{e})^{2}-2e\varphi \varepsilon_{ii}^{e}).
\end{eqnarray}

The positive terms represent microscopic processes as a contributory factor for increasing of average energy of separated micro-cracks; the negative terms are same for decreasing energy.  The constants $h_{0}$  and $\overline{h}$ are constant sink and source for average micro-crack energy. Microscopic sense of the constant $h_{0}$ is followed; it represents that part of defectiveness of a deeper structural level which provokes activation of new micro-cracks. Following from it increasing of micro-crack density reduces the elastic property of the material and, as consequence, the average energy of a micro-crack. Constant $\overline{h}$ represents that part of defectiveness of the deeper structural level which provokes micro-crack healing, what leads to decreasing of defect density and increasing of the average energy of the remaining micro-cracks. Regeneration of defects of the deeper structural level is supported by irreversible work. It is notice that in a particular problem the constants $h_{0}$  and $\overline{h}$ may be integrated in one effective constant $\overline{h}_{0}=h_{0}-\overline{h}$.

The term $h_{1} \varphi$ also represents micro-crack healing but connected with surplus energy relaxation of the self micro-cracks (self-action). The term $h_{2} \varphi^{2}$ represents micro-crack merge. At small $\varphi$ this process is weak and it is easily compensated by the above described processes. At large $\varphi$ micro-crack density increases sharply, in the same time the spatial distribution of the micro-cracks is not totally independent, but forms chains of macroscopic length from merged micro-cracks, that is, new objects of a higher structural level. Although energy of the new object (macro-crack) is high the part of energy per one micro-crack included in it becomes less, what is represented by the minus sign before the term.

In the given approximation free energy curve is a cubic parabola (Fig. \ref{f2}). In accordance with a sign of the
\begin{figure*}
  \includegraphics [width=3.5 in] {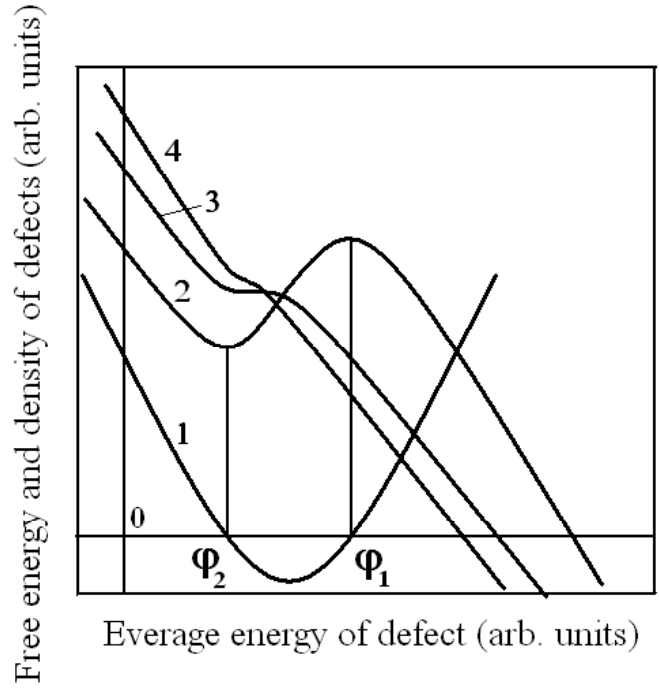}
  \caption{\label{f2}
 Schematic presentation of density of defects and free energy versus average energy of defects: 
 0 – onset of coordinate system;
 1 – equation of state $h=h(\varphi)$; 
 charts of free energy: 2 – with two extremes; 3 – with point of inflexion; 4 – declining curve}
\label{f2}
\end{figure*}
determinant
  \begin{equation}\label{a21}
D=(h_{1}+2e\varepsilon_{ii}^{e})^{2})+4h_{2}(\overline{h}_{0}-\frac{1}{2}\overline{\lambda}(\varepsilon_{ii}^{e})^{2}-\overline{\mu}(\varepsilon_{ij}^{e})^{2})
  \end{equation}
three kinds of cubic parabola are possible: with two extremes at $D>0$ (curve 2), monotonic with point of inflexion at $D=0$ (curve 3), and monotonic without point of inflexion at $D<0$ (curve 4). Which one of the three kinds of cubic parabola will be realized in a particular case depends on the relation of material parameters and the magnitude and character of a stressed state (value of tensor $\varepsilon_{ij}^{e}$)

The equation of state calculated according to (\ref{a19}) is represented by the curve 1 in Fig. \ref{f2}. It is seen that at such consideration, the extremes of free energy correspond to zero values of defect density, i.e. from the point of view of mechanics, to a continuous defect-free solid. Defect density between extremes has a negative value, and it seemingly has no physical meaning. Because of this there is something like a gap between the right and left branches of the free energy, which prohibits a spontaneous transition from one branch to another. The plot of the state equation for the free energy described by the curve 3 just touches the abscissa axis and for that described by the curve 4 it is above the abscissa axis. In both cases the defect density decreases at first and then increases with increase in the average defect energy. It implies that due to internal processes the material in the given stressed state (at predetermined $\varepsilon_{ij}^{e}$) first tends to be consolidated and only then loosened and collapsed. Presence of the free energy gap results in that the transition from the consolidation stage to the destruction one cannot be realized exclusively through the internal processes. In this case, material is stable and will not be destroyed during unlimited time.

The left extreme of the free energy is the minimum and, corresponding to it, state is stable. The right extreme is the maximum and, corresponding to it, state is unstable and provokes destruction of a material. The free energy difference of these two states represents an energy barrier. Existence of such barrier is a prerequisite for sustained stability of a material. At large stress the state is described by the free energy in the forms 3, 4. In these cases, the energy barrier vanishes, and material will be in a state of long-time creep-like destruction that will finally result in its microscopic destruction.

\subsection{Numerical calculations}

Let us assume for numerical estimates the following set of reference model parameters: $f_{0}=2\cdot10^{8} Jm^{-3}$, $\overline{h}_{0}=10^{9}m^{-3}$, $h_{1}=1.1\cdot10^{12}J^{-1}m^{-3}$, $h_{2}=0.35\cdot10^{-14} J^{-2}m^{-3}$, $\lambda=\mu=0.208\cdot10^{11} Pa$, $g=1.5\cdot10^{11} m^{-3}$, $\overline{\lambda}=0$, $\overline{\mu}=3.5\cdot10^{14} m^{-3}$, $e=5.6\cdot10^{10} J^{-1}m^{-3}$. In all calculations, the main components of the deformation tensor are assumed to be constant and equal to $\varepsilon_{11}^{e}=\varepsilon_{22}^{e}=\varepsilon_{33}^{e}=0.001$ that corresponds to uniform tension.

Changes of a form and height of potential barrier with variations of the parameters are illustrated in Fig. \ref{f3}. The decrease in the parameter $\overline{h}_{0}$ in comparison with the reference value increases the height of the potential barrier (curve 1, Fig. \ref{f3}a) 
\begin{figure*}
  \includegraphics [width=2.0 in] {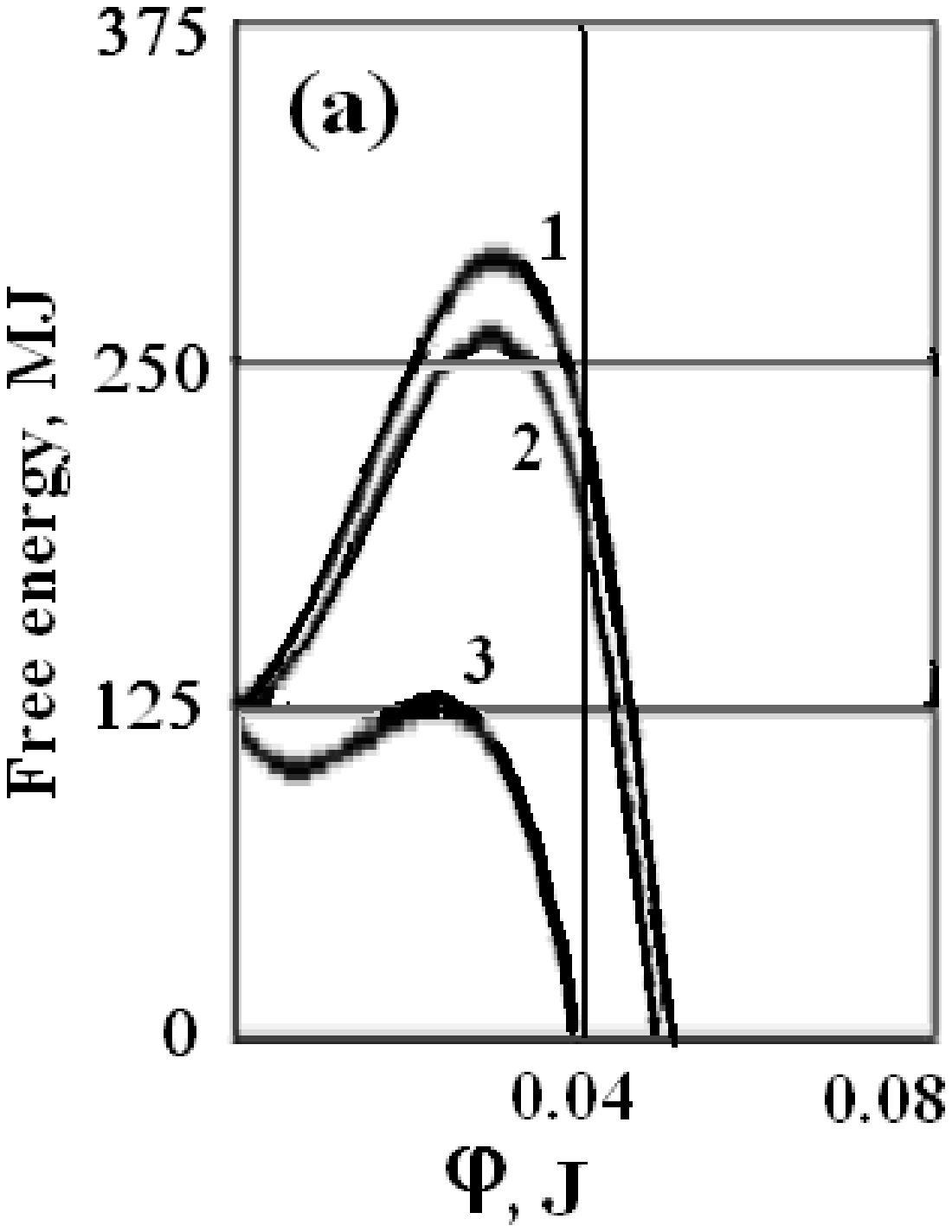}
  \includegraphics [width=2.0 in] {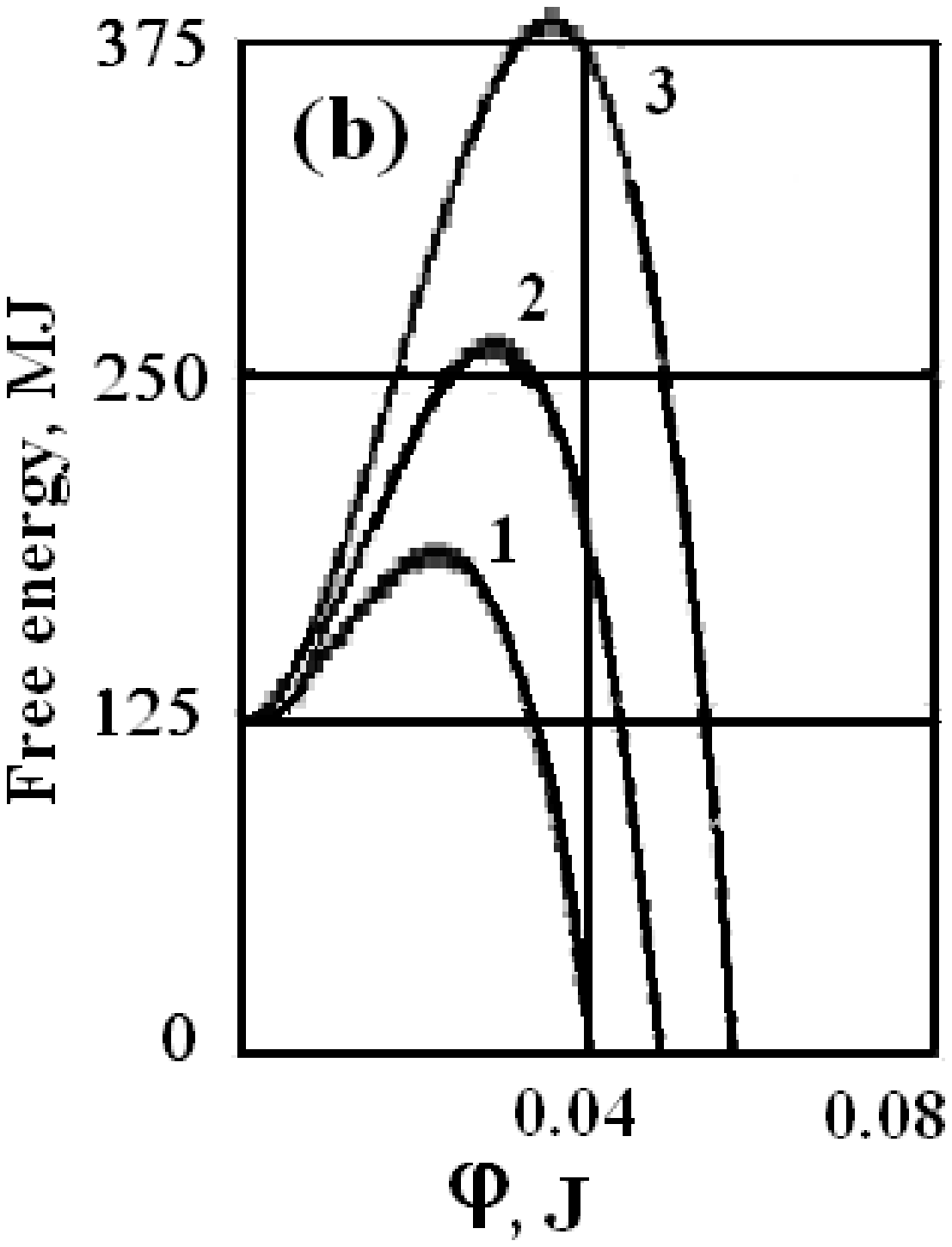}
  \includegraphics [width=2.0 in] {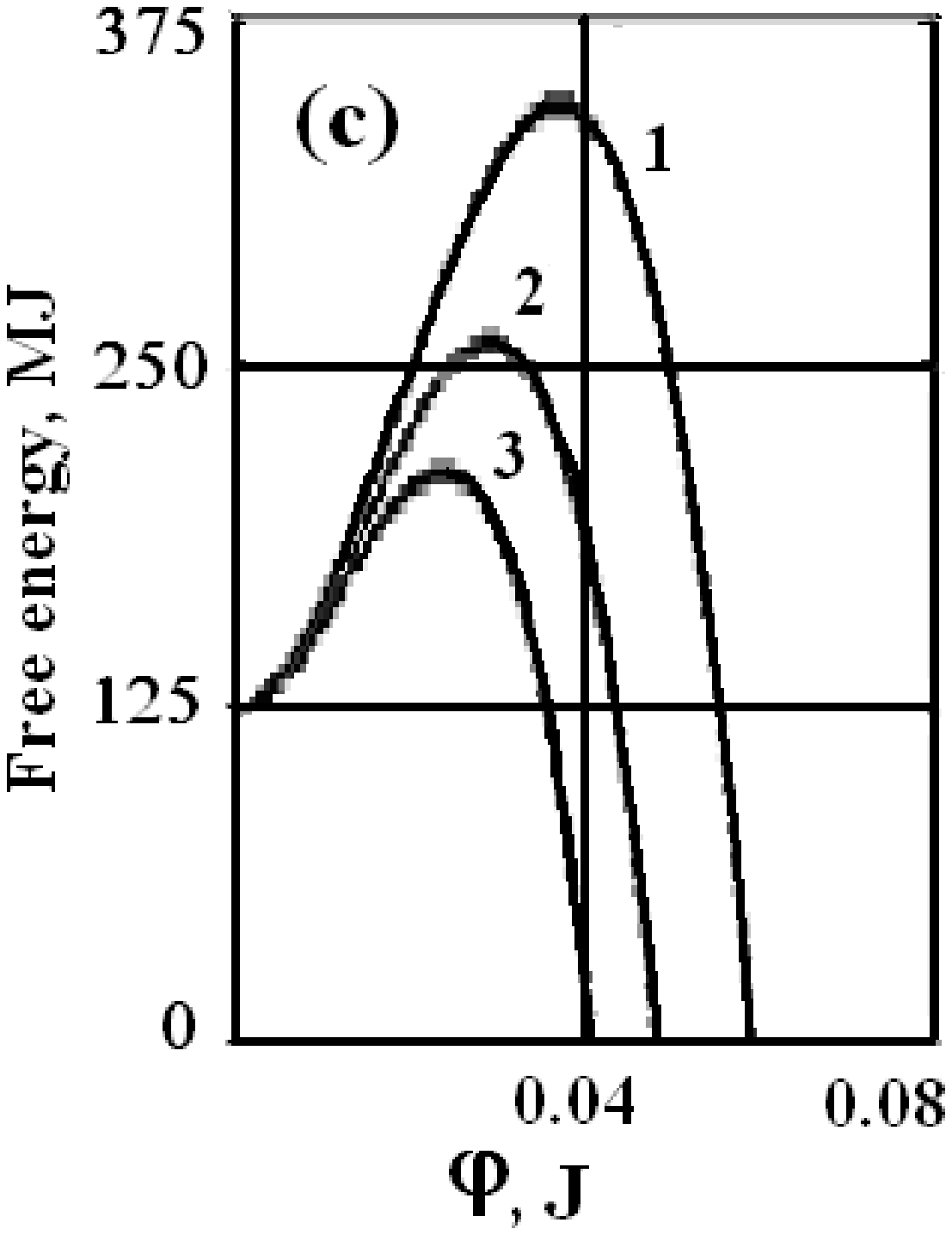}
  \caption{\label{f3}
Dependence of free energy versus average energy per microcrack with variations of parameters:
  a - $\overline{h}_{0}$,
  b - $h_{1}$,
  c - $h_{2}$.
Numbers mean the following: 2 – reference parameters; 1 –decrease ones; 3 – increase ones}
\label{f3}
\end{figure*}
that is equivalent to increase in material strength. This does not contradict common sense and satisfies our intuition, namely, the smaller is the damage of a material on a deeper structural level, and the more stable it is. The increase in the parameter $\overline{h}_{0}$ changes the height of the potential barrier sharper than in the decrease case (curve 3, Fig. \ref{f3}a). It is necessary to note, that the free energy minimum at large damage on a deeper structural level (curve 3, Fig. \ref{f3}a) has shifted noticeably towards large values of $\varphi$. It means that in a microscopically more damaged material the states with large average energy per a microcrack are much more preferential from the point of view of thermodynamics. Inasmuch as defects of larger sizes have larger energies, the instability of material can be connected with achievement by microcracks of some critical sizes or critical average energy.

Smaller values of the parameter $h_{1}$ (curve 1, Fig. \ref{f3}b), on the contrary, result in decrease in the potential barrier and describe materials with less strength, and its larger values represent materials with higher strength (curve 3, Fig. \ref{f3}b). This can be explained by the fact that decrease in efficient parameters of material with increase in the parameter $h_{1}$ that taking into account interaction of microcracks leads to the growth of adaptation capabilities of a material, the reduction of internal stress concentration that results in growth of its stability.

Influence of the parameter $h_{2}$ on stability is outwardly the same as that of the parameter $\overline{h}_{0}$ (compare Fig. \ref{f3}а and Fig. \ref{f3}c). Namely, if it decreases the stability raises (curve 1, Fig. \ref{f3}c) and if it growths the stability reduces (curve 3, Fig. \ref{f3}c). However, this influence is realized according to the other mechanism in comparison with case for the parameter $\overline{h}_{0}$. Namely, with growth $h_{2}$ the processes of coalescence of old microcracks and generation of new ones are stimulated to a greater extent that does result in a reduction of material strength.

The dependence on the other parameters of the free energy expansion is determined by what power of the parameter $\varphi$ they stand at. All items having the same power with respect to $\varphi$,  , can be united through the effective constants of the theory:
\begin{eqnarray}\label{a22}
\nonumber
h^{*}_{0}=\overline{h}_{0}+g\varepsilon_{ii}^{e}+\frac{1}{2}\overline{\lambda}(\varepsilon_{ii}^{e})^{2}+
\overline{\mu}(\varepsilon_{ij}^{e})^{2} ,\\
h^{*}_{1}=h_{1}+2e \varepsilon_{ii}^{e}).
\end{eqnarray}

The second term in the expression $h^{*}_{0}$ is linear with respect to the reversible deformation. At the tension $\varepsilon_{ii}^{e}>0$ stability is reduced with the growth of the reversible deformation (Fig. \ref{f3}a). At the compression, vice versa, $\varepsilon_{ii}^{e}<0$ and stability is increased with the growth of the reversible deformation that is confirmed by everyday experience. The last terms in the expression $h^{*}_{0}$ are quadratic with respect to deformation. It means that at positive parameters $\overline{\lambda}$ and $\overline{\mu}$ stability of a material is equally reduced, both as in case of the tension and as in case of the compression. Further, we will assume that $\overline{\lambda}<<\overline{\mu}$, i.e. shear deformations play a major role in the destruction of a material.

In the above examples, the deformation remained a constant value. It is necessary to clarify how a system described by the theory behaves in a thermodynamic cycle. Fig. \ref{f4} 
\begin{figure*}
  \includegraphics [width=2.0 in] {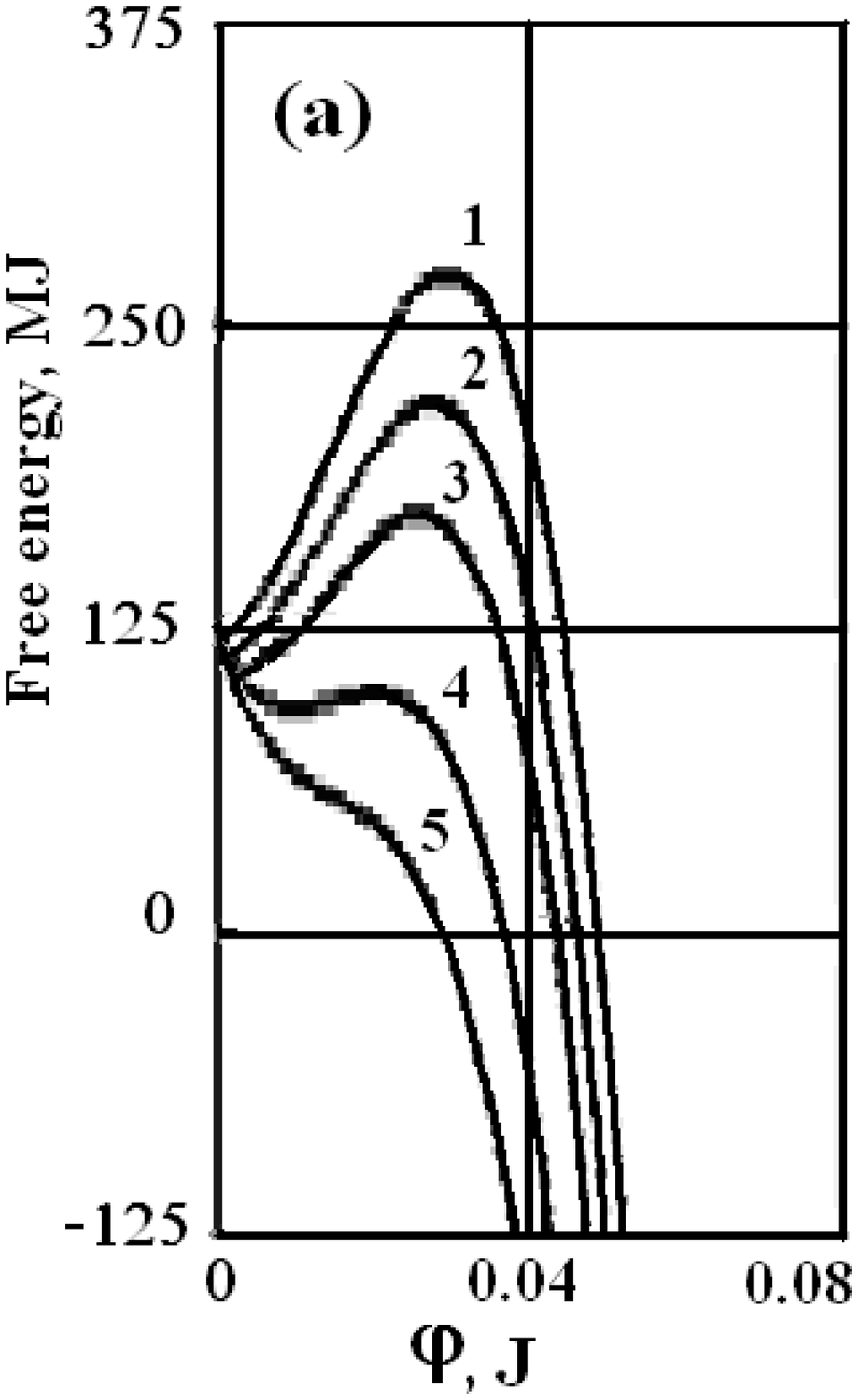}
  \includegraphics [width=2.0 in] {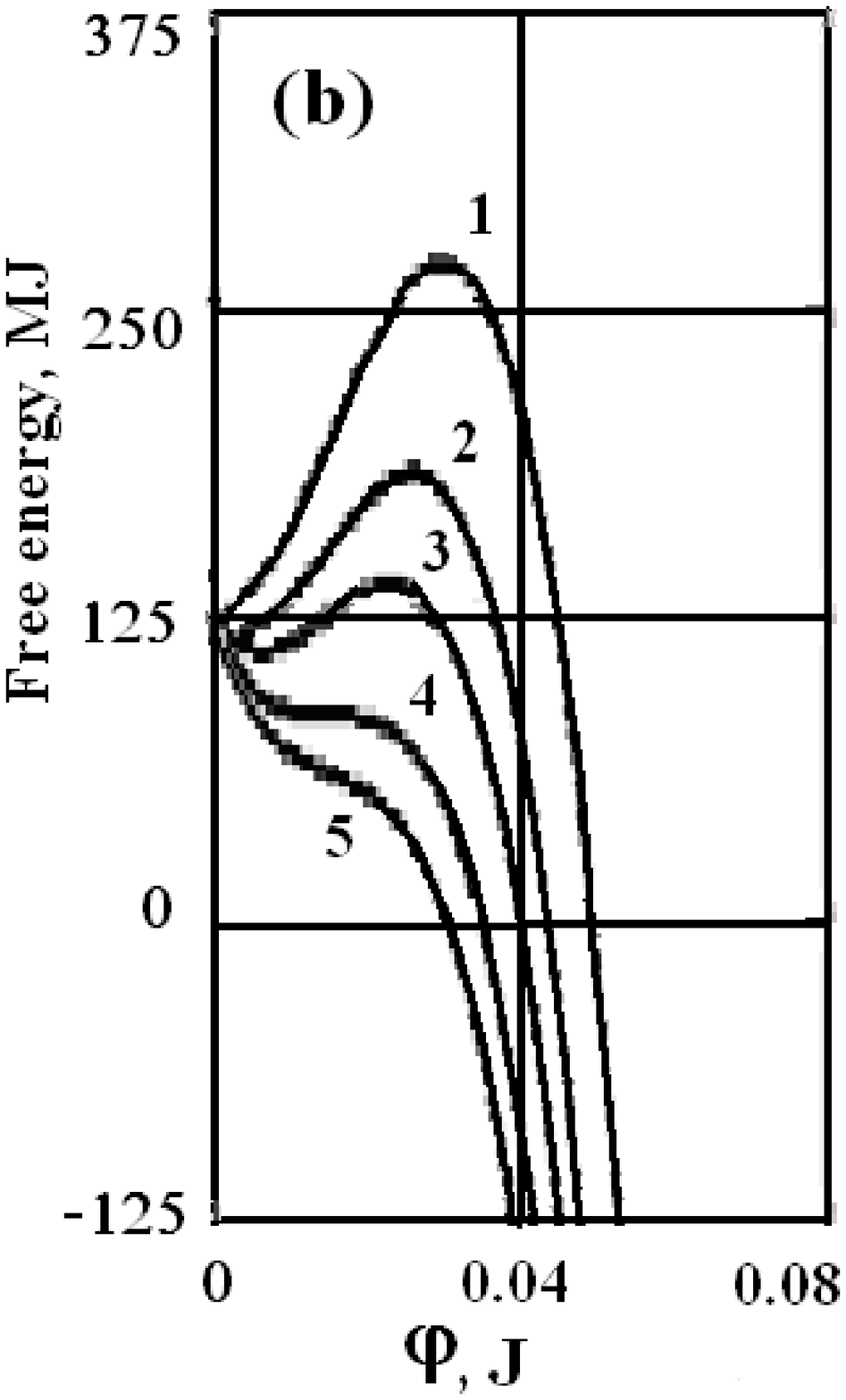}
  \caption{\label{f4}
Dependence of free energy versus average energy per microcrack during:
  a - a uniform tension;
  b - a three-axial unequal-component compression.}
\label{f4}
\end{figure*}
gives results of calculations of uniform tension and three-axial unequal-component compression. In the first case, the main components of elastic deformation tensor are equal to $\varepsilon_{11}^{e}=\varepsilon_{22}^{e}=\varepsilon_{33}^{e}=\varepsilon$ where $\varepsilon$ is the value given in Fig. \ref{f4}. In the latter case, they are equal to $\varepsilon_{11}^{e}=0$, $\varepsilon_{22}^{e}=\varepsilon$, $\varepsilon_{33}^{e}=2\varepsilon$.  Such conditions are specially chosen to describe a situation in which rocks are in the vicinity of a free surface of a mine working as in a real situation. It is seen that in both cases the height of potential barrier decreases during increase in load, and at a definite deformation it becomes equal to zero. Material loses stability both in the first and in the second case. However, instability mechanisms are different. In the first case it is due with long-term ultimate tensile strength, in the second one it is due with shear strength. However, critical values for a quasibrittle material in the given theory are not introduced explicitly as the fundamental strength constants, they are derived from the theory.

\section{EXPLANATION OF EXPERIMENTAL FACTS AND APPLIED ASPECTS OF THE PROBLEM}

The theory presented here can be applied to qualitative explanation of phenomenon of rock failure around mine workings. It is well known that at small depths mine workings have a rather high degree of stability. Around identical workings at large depths one can observe phenomena of anomalously large displacements of rocks inside the free working space that give large reduction in the mine working section area and finally lead to the complete collapse of it \cite{Shem86}. According to the theory, at small values of deformation tensor $\varepsilon_{ij}^{e}$, i.e. at small depths (estimates are given below) for rocks surrounding a mine working the condition of existence of equilibrium state is fulfilled (the case of $D<0$) which is described by free energy with one minimum (curve 2 in Fig. \ref{f2}). At large values of shear deformations (in elastic approach the volumetric deformation is equal to zero), i.e. at large depths, kind of the free energy dependence is changed, and it is already described by the curve without extremes (curve 3 in Fig. \ref{f2}a). In this case, a system has no equilibrium states (at $D>0$) and therefore is permanently collapsed.

The phenomenon described above suggests an elegant method to determine an additional estimate coupling among the parameters of the theory. Let us assume that mine depth $H$,  , starting from which said phenomenon begins to manifest itself, equals to $200m$. Corresponding to it hydrostatic pressure is $\gamma H$ and the tangential component of the stress tensor on the boundary of mine working is $2\gamma H$ (where $\gamma=\rho g$, $\rho$ is rock density equal, in the mean, to $2.6\cdot10^{3} kg\cdot m^{3}$, $g$ is the gravity constant). From here $\sigma_{\tau}\simeq 10.4 MPa$. Assuming that shear hardness of rocks $\mu=30GPa$, we get a value for the shear component of the deformation tensor of the order of $\varepsilon_{\tau}=\sigma_{\tau}{\tau} / \mu=0.00035$. A square of this value has an order of the second invariant of deformation tensor $(\varepsilon_{ij}^{e})^{2}$ that appears in free energy expansion (\ref{a18}). Taking into account that on the boundary of the mine working the first invariant of the deformation tensor is equal to zero (this follows from the exact problem solution in the elastic formulation), the zero condition for the determinant (\ref{a21}) can be written as
  \begin{equation}\label{a23}
h_{1}^{2}-4h_{2}(\overline{h}_{0}-0.00035\overline{\mu})=0.
  \end{equation}

This coupling of the theory parameters, parallel with attraction of the special experiment data, may be of use for numerical estimates. At present a complete complex of such experiments aimed at measurement of parameters $\overline{h}_{0}$, $h_{1}$, $h_{2}$, $\overline{\mu}$ and other parameters of the theory is not available. The consideration suggested here may serve as an impetus to exploratory developments in this line.

\section{CONCLUSION}

Thus, a new mesoscopic non-equilibrium thermodynamic approach is developed. The approach is based on the thermodynamic identity associated the first and second law of thermodynamics and having a perfect differential form. The introduction into the basic relation (\ref{a2}) of such internal parameters as average defect energies and defect densities has allowed to give the physical meaning of seeming purely mechanical problem of material destruction and to apply the standard methods closed to the equilibrium thermodynamics to the analysis of nonequilibrium processes. The identity permits to introduce an extended non-equilibrium state and use the good developed mathematical formalism of equilibrium and non-equilibrium thermodynamics.

The evolution of non-equilibrium variables are described by a Landau-based equation set expressed through internal or different kinds of free energy connected by means of the Legendre transforms. The introduction of some structural viscosity-like shifts $\overline{\varphi}_{k}$ and $\overline{h}_{k}$ into the evolution equations (\ref{a5}) and (\ref{a8}) make possible the accordance between the different kinds of energy.

Taking into account only one type of defects (viz., microcracks) the description of destruction of quasi-brittle solids is considered by means of an expansion of free energy. Different powers in the expansions describe thermodynamic processes of different levels. The main process is described by the lowest power; the processes corrected to it are described by higher powers of the expansions. Alternation of the signs is due to the Le Chatelier principle, namely, the processes of each next level point in opposite direction to the processes of the previous level.

The consideration allows to find stable and unstable stationary points of free energy in dependence of sign of the determinant (\ref{a21}). At $D>0$ the system has the stable stationary state, which is separated from the unstable branch by energetic gap. At $D<0$ the system has not a stable stationary state and is constantly damaged with a creep-like scenario. A large external mechanical action can change the sign of the determinant from the case $D>0$ to the case $D<0$. Owing to this change it is possible to explain the origins of the high stability of mine workings at small depths and their instability at large depths. The consideration can be used for essential enhancement of mine working stability by the way, as was announced in the ref. \cite{Met02}.

\bibliography{Metlov_MNT}

\begin{thebibliography}{11}
\expandafter\ifx\csname natexlab\endcsname\relax\def\natexlab#1{#1}\fi
\expandafter\ifx\csname bibnamefont\endcsname\relax
  \def\bibnamefont#1{#1}\fi
\expandafter\ifx\csname bibfnamefont\endcsname\relax
  \def\bibfnamefont#1{#1}\fi
\expandafter\ifx\csname citenamefont\endcsname\relax
  \def\citenamefont#1{#1}\fi
\expandafter\ifx\csname url\endcsname\relax
  \def\url#1{\texttt{#1}}\fi
\expandafter\ifx\csname urlprefix\endcsname\relax\def\urlprefix{URL }\fi
\providecommand{\bibinfo}[2]{#2}
\providecommand{\eprint}[2][]{\url{#2}}

\bibitem[{\citenamefont{Camacho and Ortiz}(1996)}]{CO96}
\bibinfo{author}{\bibfnamefont{G.~T.} \bibnamefont{Camacho}} \bibnamefont{and}
  \bibinfo{author}{\bibfnamefont{M.}~\bibnamefont{Ortiz}},
  \bibinfo{journal}{International Journal of Solids and Structures}
  \textbf{\bibinfo{volume}{33}}, \bibinfo{pages}{2899} (\bibinfo{year}{1996}).

\bibitem[{\citenamefont{Remmers et~al.}(2005)\citenamefont{Remmers, de~Borst,
  and Needleman}}]{RBN05}
\bibinfo{author}{\bibfnamefont{J.~J.~C.} \bibnamefont{Remmers}},
  \bibinfo{author}{\bibfnamefont{R.}~\bibnamefont{de~Borst}}, \bibnamefont{and}
  \bibinfo{author}{\bibfnamefont{A.}~\bibnamefont{Needleman}},
  \bibinfo{journal}{Proceeding of VIII International Conference on
  Computational Plasticity (COMPLAS VIII)} pp. \bibinfo{pages}{573--576}
  (\bibinfo{year}{2005}).

\bibitem[{\citenamefont{Van and Vasarhelyi}(2001)}]{VV01}
\bibinfo{author}{\bibfnamefont{P.}~\bibnamefont{Van}} \bibnamefont{and}
  \bibinfo{author}{\bibfnamefont{D.}~\bibnamefont{Vasarhelyi}},
  \bibinfo{journal}{In Rock Mechanics in the National Interest}
  \textbf{\bibinfo{volume}{1}}, \bibinfo{pages}{767} (\bibinfo{year}{2001}).

\bibitem[{\citenamefont{Mishnaevsky}(1997)}]{Mish97}
\bibinfo{author}{\bibfnamefont{L.~L.} \bibnamefont{Mishnaevsky}},
  \bibinfo{journal}{Engineering Fracture Mechanics}
  \textbf{\bibinfo{volume}{57}}, \bibinfo{pages}{625} (\bibinfo{year}{1997}).

\bibitem[{\citenamefont{Metlov and Morozov}(1997)}]{Met97}
\bibinfo{author}{\bibfnamefont{L.~S.} \bibnamefont{Metlov}} \bibnamefont{and}
  \bibinfo{author}{\bibfnamefont{A.~F.} \bibnamefont{Morozov}},
  \bibinfo{journal}{High-Pressure Physics and Technology}
  \textbf{\bibinfo{volume}{7}}, \bibinfo{pages}{58} (\bibinfo{year}{1997}),
  \bibinfo{note}{in Russian}.

\bibitem[{\citenamefont{Metlov}(2002)}]{Met02}
\bibinfo{author}{\bibfnamefont{L.~S.} \bibnamefont{Metlov}},
  \bibinfo{journal}{http://arxiv.org/abs/cond-mat/0204361}
  p.~\bibinfo{pages}{1} (\bibinfo{year}{2002}).

\bibitem[{\citenamefont{Metlov and Antsiferov}(2007)}]{Met07}
\bibinfo{author}{\bibfnamefont{L.~S.} \bibnamefont{Metlov}} \bibnamefont{and}
  \bibinfo{author}{\bibfnamefont{A.~V.} \bibnamefont{Antsiferov}},
  \bibinfo{journal}{High-Pressure Physics and Technology}
  \textbf{\bibinfo{volume}{17}}, \bibinfo{pages}{20} (\bibinfo{year}{2007}),
  \bibinfo{note}{in Russian}.

\bibitem[{\citenamefont{Kerner and Osipov}(1994)}]{KO94}
\bibinfo{author}{\bibfnamefont{B.~S.} \bibnamefont{Kerner}} \bibnamefont{and}
  \bibinfo{author}{\bibfnamefont{V.~V.} \bibnamefont{Osipov}},
  \emph{\bibinfo{title}{Autosolitons: A New Approach to Problems of
  Self-Organization and Turbulence}} (\bibinfo{publisher}{Kluwer},
  \bibinfo{address}{Dordrecht}, \bibinfo{year}{1994}).

\bibitem[{\citenamefont{Shemjakin et~al.}(1986)\citenamefont{Shemjakin,
  Fisenko, Kurlenja, Oparin, and another}}]{Shem86}
\bibinfo{author}{\bibfnamefont{E.~I.} \bibnamefont{Shemjakin}},
  \bibinfo{author}{\bibfnamefont{G.~L.} \bibnamefont{Fisenko}},
  \bibinfo{author}{\bibfnamefont{M.~V.} \bibnamefont{Kurlenja}},
  \bibinfo{author}{\bibfnamefont{V.~N.} \bibnamefont{Oparin}},
  \bibnamefont{and} \bibinfo{author}{\bibnamefont{another}},
  \bibinfo{journal}{Physical and Technical Problems of Mineral Wealth Mining}
  p.~\bibinfo{pages}{3} (\bibinfo{year}{1986}), \bibinfo{note}{in Russian}.

\bibitem[{\citenamefont{Metlov et~al.}(2002)\citenamefont{Metlov, Morozov, and
  Zborshchik}}]{MMZ02}
\bibinfo{author}{\bibfnamefont{L.~S.} \bibnamefont{Metlov}},
  \bibinfo{author}{\bibfnamefont{A.~F.} \bibnamefont{Morozov}},
  \bibnamefont{and} \bibinfo{author}{\bibfnamefont{M.~P.}
  \bibnamefont{Zborshchik}}, \bibinfo{journal}{Journal of Mining Science}
  \textbf{\bibinfo{volume}{38}}, \bibinfo{pages}{150} (\bibinfo{year}{2002}),
  \bibinfo{note}{in Russian}.

\bibitem[{\citenamefont{Lifshits and Pitaevskii}(1986)}]{LP86}
\bibinfo{author}{\bibfnamefont{E.~M.} \bibnamefont{Lifshits}} \bibnamefont{and}
  \bibinfo{author}{\bibfnamefont{L.~P.} \bibnamefont{Pitaevskii}},
  \emph{\bibinfo{title}{Phizicheskaia Kinetika (Physical Kinetics)}}
  (\bibinfo{publisher}{Nauka}, \bibinfo{address}{Moscow},
  \bibinfo{year}{1986}), \bibinfo{note}{in Russian}.

\end{thebibliography}

\end{document}